\pdfoutput=1
\documentclass[11pt]{article}

\usepackage[final]{acl}

\usepackage{times}
\usepackage{latexsym}

\usepackage[T1]{fontenc}

\usepackage[utf8]{inputenc}

\usepackage{microtype}

\usepackage{inconsolata}

\usepackage{bm}
\usepackage{tikzducks}
\usepackage{graphicx}
\usepackage[most]{tcolorbox}
\usepackage{colortbl}
\usepackage{pifont}
\usepackage{enumitem}
\definecolor{Gray}{gray}{0.9}

\usepackage{listings}
\definecolor{codeblue}{rgb}{0.25,0.5,0.5}
\definecolor{codekw}{rgb}{0.85, 0.18, 0.50}
\lstdefinestyle{mystyle}{
    backgroundcolor=\color{white},
    basicstyle=\fontsize{7.5pt}{7.5pt}\ttfamily\selectfont,
    columns=fullflexible,
    breaklines=true,
    captionpos=b,
    commentstyle=\fontsize{7.5pt}{7.5pt}\color{codeblue},
    keywordstyle=\fontsize{7.5pt}{7.5pt}\color{codekw},
}
\usepackage{graphicx}
\usepackage{amsmath, amssymb, bm}
\usepackage[ruled,vlined]{algorithm2e}
\usepackage{algpseudocode}
\usepackage{multirow}
\usepackage{graphicx}
\usepackage{booktabs}
\usepackage{multicol}
\usepackage{hyperref}
\usepackage{amsmath}
\usepackage{caption}
\usepackage{enumitem}
\usepackage{wrapfig}
\usepackage{listings}
\usepackage{pifont}
\usepackage{array}
\usepackage{placeins}
\usepackage{float}
\usepackage{tikz}
\newcommand*\circled[1]{\tikz[baseline=(char.base)]{
            \node[shape=circle,draw,inner sep=0.4pt] (char) {#1};}}
\definecolor{codegreen}{rgb}{0.0, 0.411, 0.243}
\definecolor{codered}{rgb}{0.89, 0.26, 0.20}
\usepackage{hyperref}
\definecolor{dartgreen}{HTML}{00693e}
\definecolor{refcolor}{HTML}{9F363A}

\hypersetup{
    colorlinks=true,
    linkcolor=refcolor,
    citecolor=dartgreen,
    filecolor=magenta,      
    urlcolor=refcolor,
    }

\title{Learning Sparsity for Effective and Efficient Music Performance \\ Question Answering}
\author{
 \textbf{Xingjian Diao}$^{1}$,
 \textbf{Tianzhen Yang}$^{2}$,
 \textbf{Chunhui Zhang}$^{1}$,
  \\
 \textbf{Weiyi Wu}$^{1}$\textbf{,}
 \textbf{Ming Cheng}$^{1}$\textbf{,}
 \textbf{Jiang Gui}$^{1}$
\\
 $^{1}$Dartmouth College, $^{2}$Yale University
 \\
   \texttt{xingjian.diao.gr@dartmouth.edu}
 \\
}
\begin{document}
\maketitle
\begin{abstract} 
Music performances, characterized by dense and continuous audio as well as seamless audio-visual integration, present unique challenges for multimodal scene understanding and reasoning. Recent Music Performance Audio-Visual Question Answering (Music AVQA) datasets have been proposed to reflect these challenges, highlighting the continued need for more effective integration of audio-visual representations in complex question answering. However, existing Music AVQA methods often rely on dense and unoptimized representations, leading to inefficiencies in the isolation of key information, the reduction of redundancy, and the prioritization of critical samples. To address these challenges, we introduce \texttt{Sparsify}, a sparse learning framework specifically designed for Music AVQA. It integrates three sparsification strategies into an end-to-end pipeline and achieves state-of-the-art performance on the Music AVQA datasets. In addition, it reduces training time by 28.32\% compared to its fully trained dense counterpart while maintaining accuracy, demonstrating clear efficiency gains. To further improve data efficiency, we propose a key-subset selection algorithm that selects and uses approximately 25\% of MUSIC-AVQA v2.0 training data and retains 70–80\% of full-data performance across models.
\end{abstract}

\section{Introduction}
\label{sec:intro}
\begin{figure}[tb]
\centering
\resizebox{0.482\textwidth}{!}{
\includegraphics{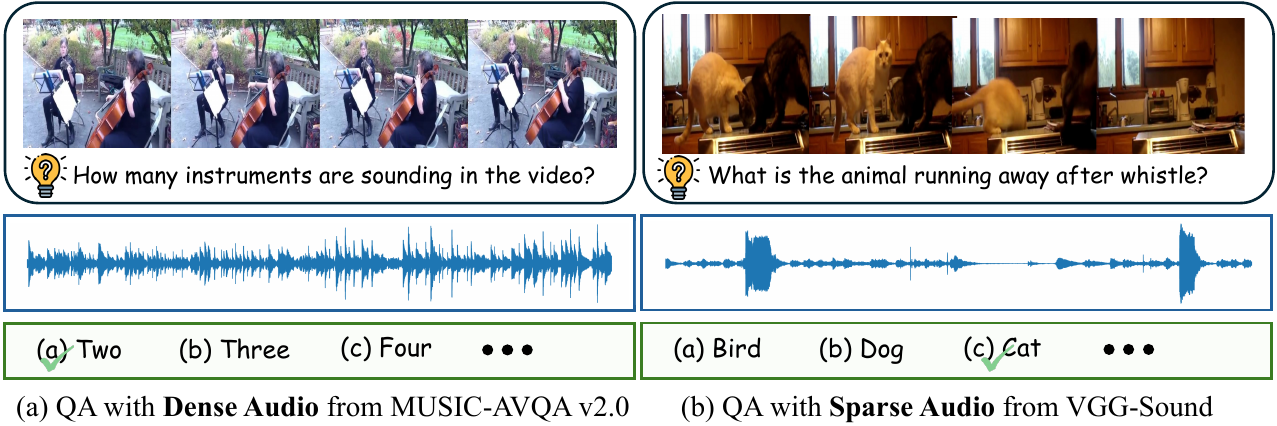}
}
\caption{
Dense Audio QA \cite{liu2024tackling} vs. Sparse Audio QA \cite{9053174}. Music performances contain dense and continuous audio signals with substantial inherent redundancy, much of which is irrelevant to the question being asked. \textbf{Sparse learning} has the potential to effectively filter out such redundancies, enabling more efficient and accurate reasoning.}
\label{fig:teaser}
\vspace{-0.7cm}
\end{figure}

Music performances, with their dense, continuous audio and seamless audio-visual integration, present both challenges and opportunities for multimodal scene understanding and reasoning \cite{you2025Music,diaoamuse}. To address the complexities of audio-visual reasoning in music scenarios, the task of Music Performance Audio-Visual Question Answering (Music AVQA) has been proposed, and the corresponding datasets, MUSIC-AVQA \cite{li2022learning} and its extended version MUSIC-AVQA v2.0 \cite{liu2024tackling}, with an example shown in Figure~\ref{fig:teaser}(a), have been introduced to facilitate research in this emerging area.

Existing AVQA methods for music performances have evolved from early cross-modality learning approaches developed for speech recognition \cite{ngiam2011multimodal, srivastava2012multimodal} to more recent advancements in multimodal fusion \cite{yun2021pano, yang2022avqa}, positive-negative pair construction \cite{li2022learning}, and state-of-the-art models such as LAVisH \cite{lin2023vision}, which adapts pretrained ViTs for cross-modal learning, and DG-SCT \cite{duan2024cross}, which employs audio-visual prompts within frozen encoders to enhance reasoning. However, current Music AVQA methods face significant limitations in effectively modeling sparse representations, which are crucial for addressing the unique challenges posed by Music AVQA tasks. These limitations include: 
\circled{1} an overreliance on dense, unoptimized representations that struggle to isolate key information from dense audio-visual signals \cite{ye2024answering, diaoamuse};
\circled{2} a lack of effective redundancy reduction mechanisms, resulting in inefficiencies during feature extraction and model inference \cite{shang2024llava};
\circled{3} the absence of prioritization strategies to identify task-critical samples, which limits scalability and prolongs training times \cite{qin2023infobatch, li2023scaling}.

\begin{figure*}[htb]
\begin{center}
\includegraphics[width=1\linewidth]{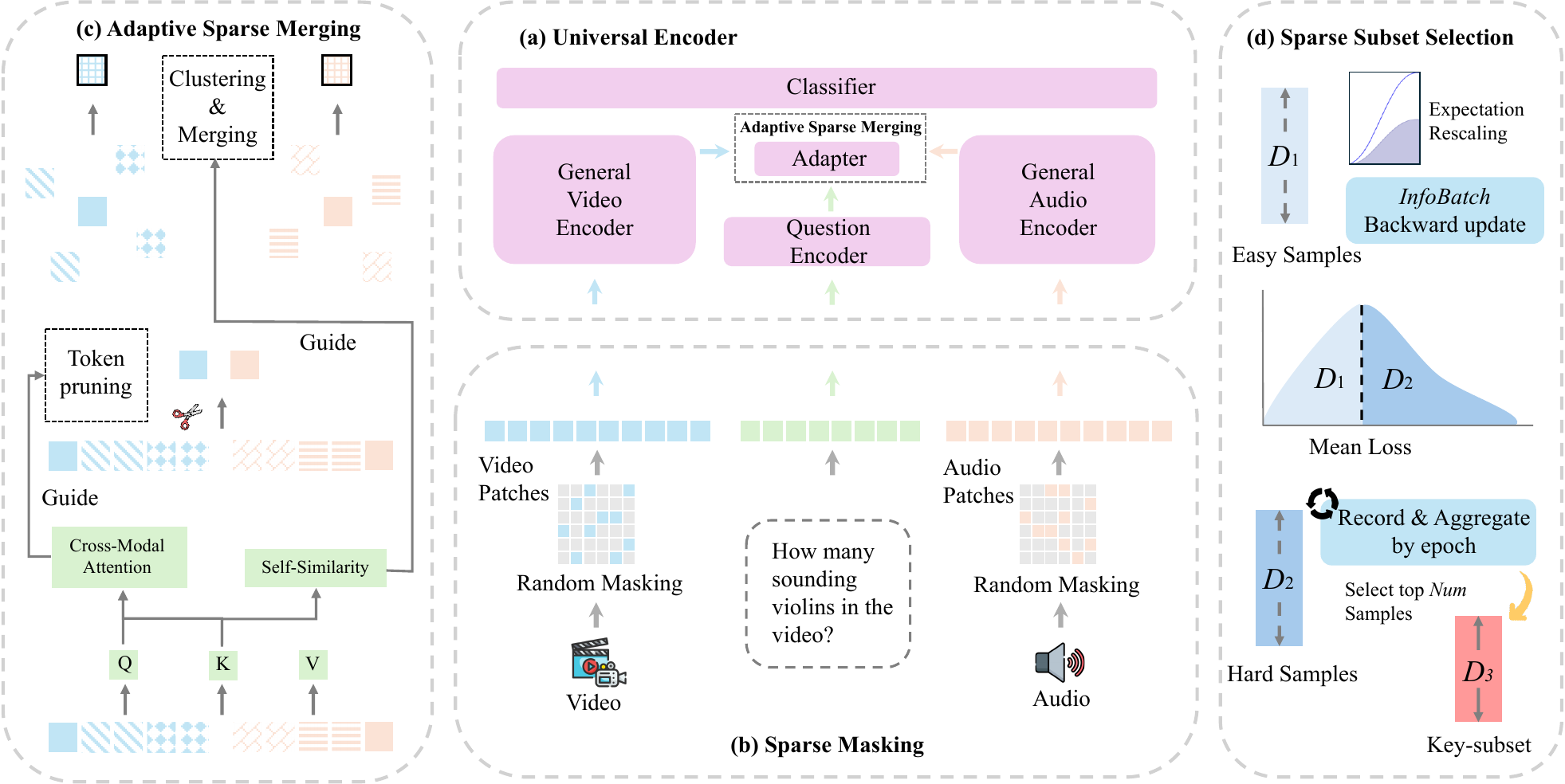}
\end{center}
\vspace{-0.3cm}
\caption{Overview of the \texttt{Sparsify} framework. \texttt{Sparsify} integrates a (a) Universal Encoder and three sparsification components: (b) Sparse Masking to reduce redundancy by masking audio and visual tokens; (c) Adaptive Sparse Merging to select and merge key multimodal tokens based on similarity; and (d) Sparse Subset Selection to prioritize impactful samples and reweight gradients with $InfoBatch$ \cite{qin2023infobatch}.}
\label{fig:model}
\end{figure*}

To address these limitations, we propose \texttt{Sparsify}, a sparse learning framework designed for Music AVQA tasks. Our contributions are:

\begin{itemize}[leftmargin=*]
\item We present an end-to-end pipeline for Music AVQA that integrates three sparsification strategies and demonstrate its effectiveness with state-of-the-art results on the MUSIC-AVQA datasets.

\item \texttt{Sparsify} reduces training time by 28.32\% while maintaining the accuracy of question answering compared to fully trained dense models, demonstrating notable efficiency improvements.

\item We introduce a key-subset selection algorithm that reduces the training dataset size by approximately 75\%, while retaining about 70-80\% of the original performance across AVQA models. 

\end{itemize}

\section{\texttt{Sparsify} Framework}
\label{sec:methods}
\subsection{Learning Multimodal Representations}
\texttt{Sparsify} adapts the Amuse framework \cite{diaoamuse} as its Universal Encoder, which includes a General Video Encoder built on Swin-V2 \cite{liu2022swin}, a General Audio Encoder leveraging the HTS-Audio Transformer \cite{chen2022hts}, and a Question Encoder based on a standard language transformer \cite{NIPS2017_3f5ee243}, as shown in Figure~\ref{fig:model}(a). Cross-modal attention is applied to align features across modalities, followed by activation and linear transformation layers, resulting in unified and informative multimodal representations tailored to Music AVQA tasks.

\subsection{Sparse Masking for Redundancy Reduction}
Music performance data inherently contain substantial redundancies, which pose significant challenges to efficient multimodal learning. Sparse Masking addresses this issue by enforcing structured sparsity to reduce redundancy and enhance computational efficiency. As illustrated in Figure~\ref{fig:model}(b), the method draws inspiration from recent advances in random masking for multimodal models \cite{li2023scaling}. This approach aligns with the objectives of sparse learning, aiming to improve efficiency while preserving model performance.

In the visual modality, Sparse Masking is applied to randomly mask 50\% of the image patches, reducing input redundancy and introducing structured sparsity to encourage more efficient visual encoding. For the audio modality, we first convert raw waveforms into mel-spectrograms and apply the same masking ratio to ensure comparable sparsity levels. This unified masking design supports coherent sparsity across modalities through consistent information reduction, contributing to more effective multimodal representation learning.

\subsection{Sparse Merging for Token Consolidation} 
In music performance AVQA tasks, dense multimodal inputs often include redundant tokens that unnecessarily increase computational overhead while diluting critical task-relevant information.  To tackle this challenge, we adapt the PruMerge method proposed by~\citet{shang2024llava}, applying it to the audio-visual setting as illustrated in Figure~\ref{fig:model}(c). This strategy dynamically prioritizes and consolidates tokens based on their significance, aligning with the objectives of sparse learning to enhance efficiency and preserve meaningful representations. Following~\citet{shang2024llava}, our approach evaluates token importance using cross-modal attention scores $\mathbf{a} = \text{softmax}\left(\frac{\mathbf{Q} \cdot \mathbf{K}^\mathbf{T}}{\sqrt{d}}\right)\mathbf{V}$, where query (\( \mathbf{Q} \)), key (\( \mathbf{K} \)), and value (\( \mathbf{V} \)) interactions highlight the relevance of each token. We then apply the Interquartile Range (IQR) method to these scores, dynamically identifying tokens within the top quartile of importance. IQR is particularly effective for filtering noise and ensuring robustness in token prioritization by focusing on outliers that represent highly salient features. Once the key tokens are identified, remaining tokens are clustered and adaptively merged with the closest key tokens according to their similarity, calculated as $\text{Similarity}(\mathbf{tok}_i,\mathbf{tok}_j) = \mathbf{k}_i \cdot \mathbf{k}_j^{T}$. This merging process retains critical tokens while integrating complementary features, preserving representational integrity. Sparse Merging ensures efficient multimodal integration with aligned sparsity across audio and visual modalities.

\subsection{Sparse Subset Selection for Efficient Training}
\label{sec:key-sub}
Training on dense audio-visual datasets is computationally expensive due to excessive redundancy. Sparse Subset Selection, illustrated in Figure \ref{fig:model} (d), addresses this by identifying and focusing on a key subset of samples that contribute the most to learning, significantly reducing training costs while preserving performance.

Following~\citet{qin2023infobatch}, our method divides samples into "hard-to-learn" (\( D_1 \)) and "easy-to-learn" (\( D_2 \)) categories based on their loss values relative to the mean. Hard samples (\( D_1 \)) are recorded and aggregated by epoch, with their importance weighted by a decay ratio \( r \) over \( k \)-epoch intervals. This ensures that difficult samples are prioritized early in training, while less critical samples are deprioritized over time. The top \( num \) samples with the highest aggregated scores are selected to form the final Key-subset (\( D_3 \)). $InfoBatch$ \cite{qin2023infobatch} is used to rescale gradients, pruning redundant "easy-to-learn" samples (\( D_2 \)) and ensuring that the reduced dataset retains the statistical properties of the original. This combination minimizes redundancy, accelerates convergence, and maintains task performance. The detailed Key-subset Selection Algorithm is presented as Algorithm~\ref{alg:key_subset_selection}.

\newcommand{\graycomment}[1]{\textcolor{gray}{\textit{\# #1}}}
\SetAlCapFnt{\small} 
\SetAlCapNameFnt{\small} 
{\small
\begin{algorithm}
\small
\caption{Key-subset Selection Algorithm}
\label{alg:key_subset_selection}
\KwIn{Model $M$, Dataset $\mathcal{D} = \{(x_i, y_i)\}_{i=1}^N$ with $N$ samples, Loss function $L$, Number of epochs $E$, Merge group size $k$, Decrement ratio $r$, Number of key samples $n$}
\KwOut{Key-subset indices $\mathcal{K}$}

\graycomment{Initialization} \\
Initialize scores vector $\mathbf{s} \gets \mathbf{1} \in \mathbb{R}^N$ \\
Initialize epochs list $\text{EpochsList} \gets [\;]$ \\

\graycomment{Compute original losses} \\
\For{$i \gets 1$ \KwTo $N$}{
    Compute loss $l_i \gets L(M(x_i), y_i)$ \\
    Update score $s_i \gets l_i$
}

\graycomment{InfoBatch} \\
\For{$\text{epoch} \; e \gets 1$ \KwTo $E$}{
    Initialize temporary count vector $\mathbf{t} \gets \mathbf{0} \in \mathbb{R}^N$ \\
    Compute mean loss $\mu \gets \frac{1}{N} \sum_{i=1}^N s_i$ \\
    \For{$i \gets 1$ \KwTo $N$}{
        Compute loss $l_i \gets L(M(x_i), y_i)$ \\
        Update score $s_i \gets l_i$ \\
        \If{$s_i > \mu$}{
            Increment count $t_i \gets t_i + 1$
        }
    }
    Append $\mathbf{t}$ to $\text{EpochsList}$
}

\graycomment{Merge} \\
Initialize merged scores $\mathbf{m} \gets \mathbf{0} \in \mathbb{R}^N$ \\
Compute number of groups $G \gets \lceil \frac{E}{k} \rceil$ \\
\For{$\text{group} \; g \gets 1$ \KwTo $G$}{
    Compute group weight $w_g \gets r^{g-1}$ \\
    \For{$\text{epoch} \; e \gets (g-1) \cdot k + 1$ \KwTo $\min(g \cdot k, E)$}{
        Update merged scores $\mathbf{m} \gets \mathbf{m} + w_g \cdot \text{EpochsList}[e]$
    }
}

\graycomment{Select the top $n$ indices as the Key-subset} \\
$\mathcal{K} \gets \text{argsort}(-\mathbf{m})[:n]$ \\
\Return Key-subset indices $\mathcal{K}$
\end{algorithm}
}

\begin{table*}[htbp]
\centering
\resizebox{2.0\columnwidth}{!}{%
\begin{tabular}{@{}l|ccc|ccc|cccccc|c@{}}
\toprule
\multirow{2}{*}{{Method}} & \multicolumn{3}{c|}{{Audio-related QA}}              & \multicolumn{3}{c|}{{Visual-related QA}}             & \multicolumn{6}{c|}{{Audio\&Visual-related QA}}                                                          & \multirow{2}{*}{{Avg}} \\
                           & {Count} & {Comp}  & {Avg}   & {Count} & {Local} & {Avg}   & {Exist} & {Count} & {Local} & {Comp}  & {Temp}  & {Avg}   &                               \\ \midrule
\multicolumn{14}{c}{MUSIC-AVQA} \\ \midrule
AVST \cite{li2022learning}  & 77.78          & 67.17          & 73.87          & 73.52          & 75.27          & 74.40          & \textbf{82.49}          & 69.88          & 64.24          & 64.67          & 65.82          & 69.53          & 71.59                         \\
LAVisH   \cite{lin2023vision}      & 75.59          & \textbf{84.13} & \underline{76.86}         & 77.45          & 72.91          & 76.29          & 71.91          & \underline{77.52}          & \underline{75.81}          & \underline{76.75}          & \underline{77.62}          & \underline{76.31}          & \underline{76.10}                         \\
DG-SCT \cite{duan2024cross} & \textbf{83.27}          & 64.56          & 76.34          & \underline{81.57}          & \underline{82.57}          & \underline{82.08}          & \underline{81.61}          & 72.84          & 65.91          & 64.22          & 67.48          & 70.56          & 74.62                         \\ 
\textbf{\texttt{Sparsify} (Ours)}   & \underline{83.12} & \underline{77.64} & \textbf{80.38} & \textbf{83.12} & \textbf{85.74} & \textbf{84.43} & 80.98 & \textbf{82.70} & \textbf{85.09} & \textbf{77.12} & \textbf{79.89} & \textbf{81.80}  & \textbf{81.75}                               \\ \midrule
\multicolumn{14}{c}{MUSIC-AVQA v2.0} \\ \midrule
AVST \cite{li2022learning}  & 81.38          & 61.82          & 75.20          & 78.72          & 77.29          & 78.05          & 71.63          & 68.62          & 64.39          & 64.03          & 60.29          & 65.83          & 70.83                         \\
LAVisH \cite{lin2023vision} & \textbf{83.82} & 58.19          & 75.72          & \textbf{82.81} & 81.73          & \underline{82.30} & 73.26 & \underline{73.45} & \underline{65.64} & 64.26          & 60.82          & 67.75 & 73.28 \\
DG-SCT \cite{duan2024cross} & \underline{83.13} & \underline{62.54} & \underline{76.62} & 81.61          & \underline{82.76} & 82.19          & \textbf{83.43} & 72.70          & 64.65          & \underline{64.78} & \underline{67.34} & \underline{70.38}          & \underline{74.53} \\ 
\textbf{\texttt{Sparsify} (Ours)}   & 82.16          & \textbf{76.44} & \textbf{79.30} & \underline{82.54} & \textbf{85.15} & \textbf{83.84} & \underline{80.68} & \textbf{82.80} & \textbf{84.89} & \textbf{76.92} & \textbf{80.09} & \textbf{81.08} & \textbf{81.30} \\
\bottomrule
\end{tabular}
}
\vspace{-0.2cm}
\caption{Comparison with state-of-the-art methods on the MUSIC-AVQA and MUSIC-AVQA v2.0 test set. Accuracy is reported for Audio (Counting, Comparative), Visual (Counting, Location), and Audio-Visual (Existential, Counting, Location, Comparative, Temporal) question types. Average accuracies for Audio, Visual, Audio-Visual, and Overall are also included. \textbf{Bold} marks the best results, and \underline{underlined} marks the second-best.}
\vspace{-0.3cm}
\label{tab:merged_with_centered_datasets}
\end{table*}

\section{Experiments}
\label{sec:exp}

\subsection{Setup}
\paragraph{Music AVQA Datasets} 
\underline{\textit{(i)}} MUSIC-AVQA \cite{li2022learning}: Contains 9,288 videos (150 hours) spanning 22 instruments, with 45,867 QA pairs derived from 33 templates across four categories: String, Wind, Percussion, and Keyboard. Each video includes approximately five QA pairs on average.  
\underline{\textit{(ii)}} MUSIC-AVQA v2.0 \cite{liu2024tackling}: An extension addressing data bias, introducing 1,230 additional videos and 8,100 new QA pairs.

\paragraph{Full Dataset Training Configuration} 
For the experiments described in Section~\ref{main reults exp}, Sparse Masking is applied during the first three epochs and is disabled thereafter. Adaptive Sparse Merging and $InfoBatch$ are used throughout the training. We set the masking rate to 50\% for Sparse Masking. For $InfoBatch$, the ratio is set to 0.5 and the delta to 0.875, following the setup in \cite{qin2023infobatch}.

\paragraph{Key-subset Selection Configuration} 
In the key-subset selection, we apply a one-epoch warm-up phase followed by 15 epochs of training. The hyperparameters are set as follows: $N = 15$, $k = 3$, $r = 0.618$, and $Num = 10{,}819$ (i.e., the number of QA pairs). During this stage, only $InfoBatch$ \cite{qin2023infobatch} is employed, while Sparse Masking and Adaptive Sparse Merging are disabled.

\begin{figure}[ht]
\begin{center}
\includegraphics[width=1\linewidth]{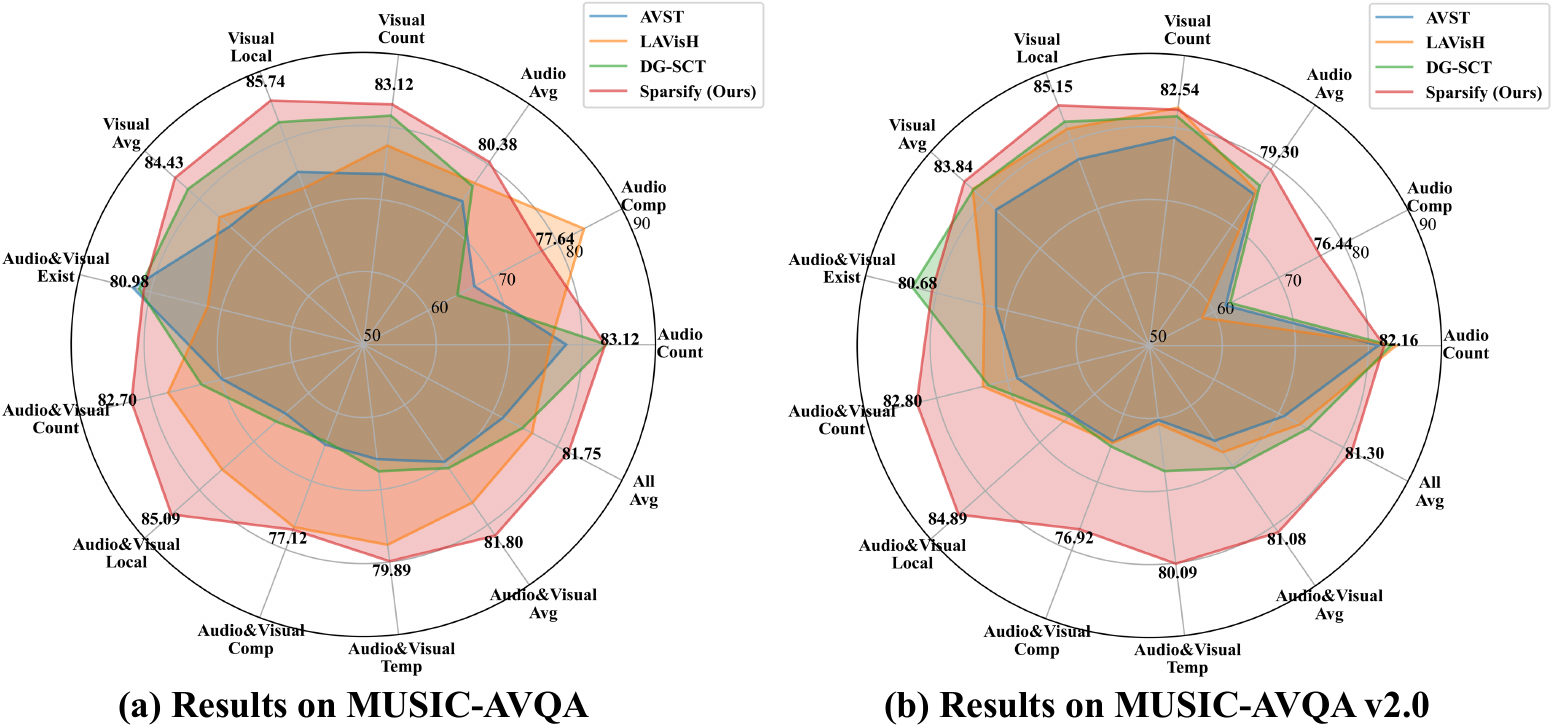}
\end{center}
\vspace{-0.3cm}
\caption{Radar charts comparing \texttt{Sparsify} with state-of-the-art methods on MUSIC-AVQA and MUSIC-AVQA v2.0, across various question types.}
\vspace{-0.4cm}
\label{fig:result_attributes}
\end{figure}

\begin{figure*}[ht]
\centering
\includegraphics[width=1\linewidth]{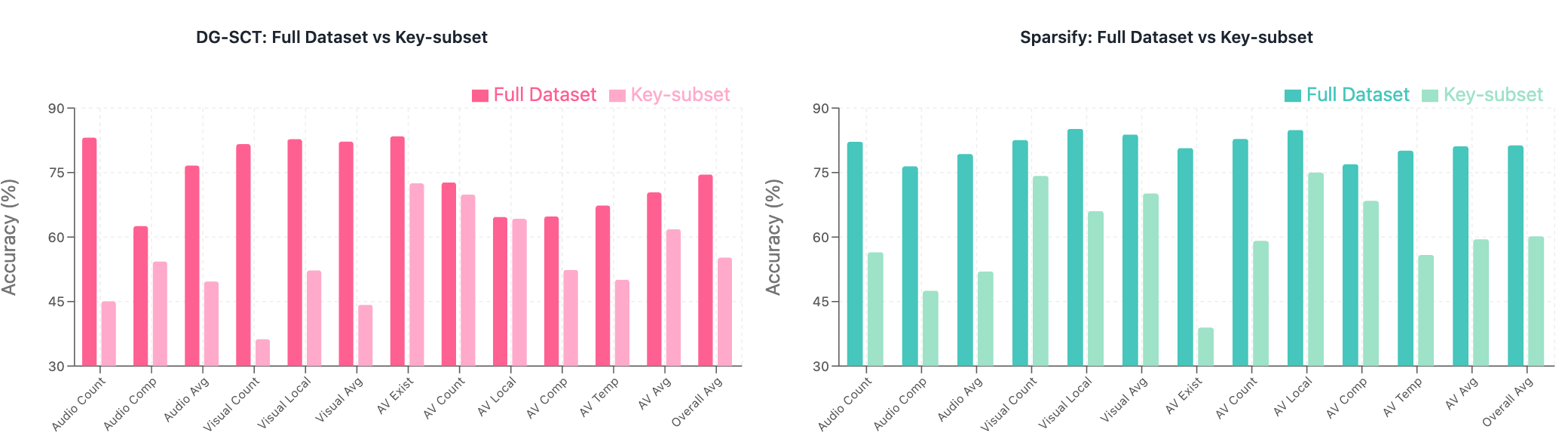}
\vspace{-0.6cm}
\caption{
Accuracy comparison of DG-SCT and \texttt{Sparsify} trained on the full dataset and the key-subset ($\sim$25\% of data) \cite{liu2024tackling}. Training on the key-subset maintains strong performance despite substantial data reduction.
}
\vspace{-0.3cm}
\label{fig:subset vs full}
\end{figure*}

\subsection{Comparison with State-of-the-Art}
\label{main reults exp}
Table~\ref{tab:merged_with_centered_datasets} presents a detailed comparison between \texttt{Sparsify} and three strong baselines—AVST~\cite{li2022learning}, LAVisH~\cite{lin2023vision}, and DG-SCT~\cite{duan2024cross}—on both the MUSIC-AVQA and MUSIC-AVQA v2.0 benchmarks. \texttt{Sparsify} achieves the highest overall average accuracy on both datasets, with consistent gains across audio, visual, and audio-visual question types. These improvements highlight the potential of sparse learning in handling the dense and continuous nature of music performance videos. 

\noindent\textbf{Sparse learning benefits visual question answering by promoting compact visual representations.}  
On visual-related QA, \texttt{Sparsify} achieves accuracy scores of 84.43\% and 83.84\% on MUSIC-AVQA and MUSIC-AVQA v2.0, respectively, outperforming DG-SCT by +2.35\% and +1.65\%, and surpassing LAVisH by +8.14\% and +1.54\%. These improvements reflect the advantage of sparse inputs in retaining essential spatial and structural cues while reducing visual redundancy. In music performance QA, such representations better support reasoning over complex scenes involving performer locations, interactions, and visual composition. By limiting the influence of background clutter and redundant details, sparse visual representation enables the model to perform more robustly across diverse and fine-grained visual reasoning types.

\noindent\textbf{Sparse learning supports audio question answering by reducing spectral redundancy and enabling efficient acoustic encoding.}
On audio-related QA, \texttt{Sparsify} achieves gains of +3.52\% and +3.58\% over LAVisH on MUSIC-AVQA and MUSIC-AVQA v2.0, respectively, and outperforms DG-SCT by +4.04\% and +2.68\%. These improvements suggest that sparse input representations help suppress redundant frequency patterns while retaining sufficient acoustic detail for question-relevant reasoning. By pruning less informative segments in the spectrogram, \texttt{Sparsify} yields more compact yet informative representations, supporting improved performance on audio-based queries involving complex musical content.

\noindent\textbf{Sparse learning improves audio-visual question answering by jointly reducing modality-specific redundancies.}  
On audio-visual QA, \texttt{Sparsify} outperforms DG-SCT by +11.24\% and +10.70\%, and LAVisH by +5.49\% and +13.33\% on MUSIC-AVQA and MUSIC-AVQA v2.0, respectively. These consistent gains suggest that sparsification across both audio and visual modalities helps eliminate less informative content and produce more streamlined multimodal representations with reduced noise. This joint reduction in redundancy allows the model to more effectively capture relevant cross-modal associations in complex performance scenarios.

\subsection{Key-subset Selection and Performance Retention}
\label{sec:Key-subset exp}

We evaluate the effectiveness of our Key-subset selection algorithm on the MUSIC-AVQA v2.0 dataset \cite{liu2024tackling}, as illustrated in Figure~\ref{fig:subset vs full}. The selected subset comprises only $\sim$25\% of the original training set (10,819 samples), yet enables models to retain a substantial portion of their full-data performance. When trained exclusively on this subset, \texttt{Sparsify} achieves 60.17\% accuracy and DG-SCT \cite{duan2024cross} achieves 55.21\%, corresponding to 74.01\% and 74.08\% of their respective accuracies when trained on the full dataset. These results demonstrate that our Key-subset Selection reduces training data usage while retaining much of the models' original performance, offering a data-efficient solution for Music AVQA.

\begin{figure}[ht]
\centering
\includegraphics[width=1\linewidth]{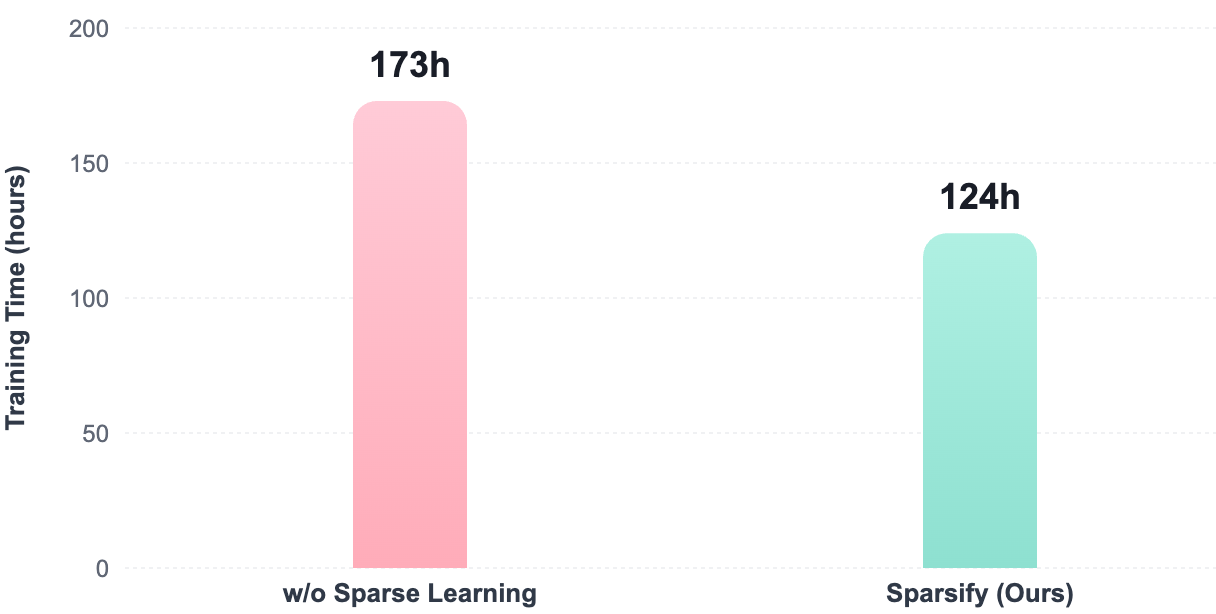}
\caption{
Comparison of the training time of \texttt{Sparsify} with a dense variant that disables all three sparsification strategies. Results are reported on the MUSIC-AVQA v2.0 dataset \cite{liu2024tackling}.
}
\vspace{-0.3cm}
\label{fig:trainingtime}
\end{figure}

\subsection{Training Efficiency Gains from Sparse Learning} 
Figure~\ref{fig:trainingtime} illustrates the training efficiency gains of \texttt{Sparsify}, which reduces total training time from 173 hours to 124 hours—a 28.32\% improvement over its dense variant. These gains reflect the combined effect of three sparsification strategies integrated into the training pipeline. Sparse Masking reduces early-stage computational load by masking 50\% of audio and visual tokens. Sparse Merging compresses intermediate representations by consolidating similar tokens, reducing token-level complexity. In parallel, using \textbf{$InfoBatch$} enhances efficiency by emphasizing harder-to-learn samples, which accelerates convergence and reduces the number of required optimization steps.

\section{Conclusion}
We present \texttt{Sparsify}, a sparse learning framework for Music AVQA that addresses the inefficiencies inherent in dense audio-visual representations. \texttt{Sparsify} achieves this by \textit{(i)} integrating three sparsification strategies into an end-to-end pipeline and achieving state-of-the-art performance on Music AVQA datasets; \textit{(ii)} reducing training time by 28.32\% while maintaining comparable accuracy to its dense counterpart. In addition, we propose a key-subset selection algorithm that selects and uses approximately 25\% of the MUSIC-AVQA v2.0 training data, while retaining 70–80\% of full-data performance across models. We hope our work offers insights into efficient multimodal understanding in dense audio-visual settings.
\clearpage
\newpage
\section*{Limitations}
The effectiveness of \texttt{Sparsify} has been demonstrated on large-scale Music AVQA benchmarks, as it is specifically designed to address the inefficiencies of dense audio-visual representations in this domain. While it may hold promise for broader multimodal tasks, its behavior in such settings remains to be explored. Extending \texttt{Sparsify} to more diverse or unconstrained applications represents a valuable direction for future work.

\section*{Ethical Considerations}
We examined the study describing the publicly available datasets used in this research and identified no ethical issues regarding the datasets. 

\section*{Acknowledgment}
This study is supported by the Department of Defense grant HT9425-23-1-0267.

\bibliography{ref}

\appendix

\section{Baselines}
\label{baselines}
\begin{itemize}[leftmargin=*]
\item \textbf{AVST} \cite{li2022learning}: 
Integrates audio, visual, and question modalities for spatio-temporal reasoning in audio-visual question answering. It aligns modalities through spatial and temporal grounding, fuses features into a joint representation, and optimizes both grounding and QA objectives.

\item \textbf{LAVisH} \cite{lin2023vision}: 
Adapts frozen Vision Transformers for audio-visual tasks using lightweight adapters and latent tokens to compress and fuse audio-visual information. Cross-modal attention and adapter modules enable bidirectional interaction between modalities.
\item \textbf{DG-SCT} \cite{duan2024cross}: Enhances audio-visual tasks through a Dual-Guided Spatial-Channel-Temporal attention mechanism, dynamically adjusting feature extraction and facilitating bidirectional audio-visual guidance with lightweight interaction layers.

\end{itemize}

\section{Related Work}
\label{sec:related_work}
\paragraph{Audio-Visual Video Understanding} 
Audio and visual modalities offer complementary cues that, when jointly modeled, support a more comprehensive understanding of the scene \cite{wei2022learning, diao2023av, shu2023audio, diao2025temporal}. 
Early work focused on joint representations for tasks like audio-visual speech recognition \cite{ngiam2011multimodal} and multimodal deep learning \cite{srivastava2012multimodal}. Recent methods enhance fusion techniques for sound source localization \cite{zhao2018sound} and audio-driven visual analysis \cite{zhao2019sound}. Frameworks such as LAVisH \cite{lin2023vision}, which proposed a latent audio-visual hybrid adapter that adapts pretrained ViTs to audio-visual tasks by injecting a small number of trainable parameters into every layer of a frozen ViT, and DG-SCT \cite{duan2024cross} which incorporates trainable cross-modal interaction layers into pre-trained audio-visual encoders, allowing adaptive extraction of crucial information from the current modality across spatial, channel, and temporal dimensions, while preserving the frozen parameters of large-scale pre-trained models. As for benchmarks, there are MUSIC-AVQA \cite{li2022learning}, AVQA \cite{yang2022avqa}, MUSIC-AVQA v2.0 \cite{liu2024tackling} and AV-Odyssey Bench \cite{gong2024av}, which focus on whether model can truly understand audio-visual information. However, existing approaches overlook the unique challenges of music performance datasets, where dense and continuous audio-visual signals lead to significant redundancy. These dense representations hinder efficient processing and dilute task-relevant features, necessitating sparsification strategies to enable efficient reasoning in this domain.

\paragraph{Multimodal Question Answering} 
Multimodal question answering spans Visual QA (VQA) \cite{antol2015vqa, lei2018tvqa, li2024towards}, Audio QA (AQA) \cite{fayek2020temporal}, and Audio-Visual QA (AVQA) \cite{li2022learning}, requiring the integration of modality-specific signals for complex reasoning tasks. For VQA, datasets such as MMMU \cite{yue2024mmmu} and MMBench \cite{liu2025mmbench} provide carefully curated benchmarks that evaluate vision-language models across diverse domains. For AQA, notable datasets include Clotho-AQA \cite{lipping2022clotho} and AIR-Bench \cite{yang2024air}, which consist of question-answering tasks derived from environmental and event-based audio scenes. AVQA benchmarks such as MUSIC-AVQA \cite{li2022learning}, AVQA \cite{yang2022avqa}, Pano-AVQA \cite{yun2021pano}, FunQA \cite{xie2024funqa}, and MUSIC-AVQA v2.0 \cite{liu2024tackling} emphasize spatio-temporal reasoning and multimodal fusion in complex video contexts. Among these, Music AVQA presents distinctive challenges due to its continuous and densely structured audio signals, making it valuable for multimodal reasoning~\cite{you2025Music}.

\paragraph{Sparse Learning in Audio-Visual Signals} 
Sparsity has become increasingly crucial in audio-visual signal processing due to the inherent complexity and redundancy of cross-modal data. Early approaches employ shift-invariant kernels \cite{monaci2008learning} to capture essential patterns while reducing computational overhead. This foundation leads to more sophisticated methods using group sparsity and joint dictionaries \cite{shen2013audio}, which are particularly effective in handling noisy and variable signals. Current research focuses on temporal dynamics in audio-visual learning, where audio-visual relationships are often intermittent but contextually meaningful \cite{iashin2022sparse}. Modern transformer-based architectures with specialized selection mechanisms \cite{iashin2024synchformer} have shown promise in processing extended sequences efficiently. However, sparsity-based approaches remain underexplored in the context of music performance question answering, where challenges such as overlapping instruments and complex audio-visual interactions demand more efficient representations. Our work aims to bridge this gap with sparsification strategies.

\section{Positioning Music AVQA Among Multimodal Tasks}
\label{music_vs_other}
To contextualize Music AVQA, it is useful to distinguish it from broader multimodal tasks that also integrate information across modalities. This section contrasts Music AVQA with vision-language modeling, audio-language modeling, and other representative domain-specific question answering to highlight its unique challenges and requirements.

\subsection{Vision-Language Modeling}

Vision-language modeling aims to enable multimodal systems to interpret visual content—such as images and videos—in conjunction with textual descriptions \cite{jian2023bootstrapping, bordes2024introduction, jian2024expedited, zhang2024vision}. It has supported a wide range of applications, including text-to-image generation \cite{li2019controllable, gao2024gem, he2025intentenhan}, video editing \cite{hu2023reinforcement, he2025enhancing}, video captioning and grounding \cite{pan2022auto, li2024towards, zhang2025pretrained}, and proxy learning \cite{qian2023intra, yao2024customized, yao2024multi}.
In contrast, Music AVQA requires integrated reasoning over continuous audio-visual streams, where visual understanding must be synchronized with rhythm, motion, and acoustic cues. This setting introduces challenges such as temporal alignment and redundancy reduction, which are not typically emphasized in standard vision-language tasks.

\subsection{Audio-Language Modeling}

Audio-language modeling \cite{borsos2023audiolm, deshmukh2023pengi, wu2024towards, su2025audio} builds systems that fuse audio features with text for various downstream tasks such as audio question answering \cite{fayek2020temporal, lipping2022clotho}, text-to-speech generation \cite{min2021meta, le2023voicebox}, and audio editing \cite{wang2023audit, liang2024wavcraft}. These tasks primarily focus on modeling relationships between acoustic signals and language, often in domains such as speech, environmental sounds, or sound events. Unlike conventional audio-language tasks that focus on modeling acoustic-linguistic associations, Music AVQA incorporates a visual modality that is intricately entangled with the audio stream in music performance contexts. This setting necessitates fine-grained multimodal reasoning, where models must jointly interpret auditory patterns, visual dynamics, and their temporal interplay to answer performance-specific questions.

\subsection{Domain-Specific Question Answering}
Domain-specific question answering systems are designed to operate within specialized fields by leveraging structured knowledge and domain-specific data. Examples such as educational, financial, and medical QA, as discussed below, entail distinct input modalities, reasoning demands, and representational challenges.

\paragraph{Educational QA} 
Educational Question Answering (Educational QA) systems \cite{soares2021education, steuer2022investigating, wu2023towards} are designed to support learning processes by responding to student queries based on educational materials such as textbooks, lecture notes, and academic articles. The primary goal is to clarify concepts, explain solutions, or guide students through subject matter. In contrast, Music AVQA involves perceptual reasoning over evolving audio-visual input. Instead of extracting explicit concepts from structured curricula, models must interpret expressive cues such as gestural nuance, visual-musical alignment, and acoustic articulation in continuous video streams. This shift demands interpreting perceptual and temporal patterns, which are not typically required in conventional educational QA. The emphasis on fluid multimodal integration further distinguishes Music AVQA as a challenging reasoning setting. 

\paragraph{Financial QA}
Financial Question Answering (Financial QA) \cite{li2023large, huang2023finbert, saini2023evolution} focuses on extracting insights and answering questions from a wide range of financial data \cite{chen2022convfinqa}, such as company reports, market data, economic indicators, and financial news. These systems assist analysts, investors, and businesses in making informed decisions by providing quick access to relevant financial information and analysis. While Financial QA can involve data from multiple views (e.g., text, tables, charts) \cite{zhu2021tat}, it typically does not involve the continuous, dense audio-visual streams found in music performances. The core task in Financial QA is to identify factual information, understand financial terminology, perform numerical reasoning, and interpret trends from often structured or semi-structured textual and numerical data \cite{wang2022novel}. In contrast, Music AVQA centers on the temporal and semantic understanding of performance events, requiring models to interpret how visual gestures correspond to musical outcomes, such as identifying sounding instruments, tracking temporal changes, and linking expressive motion to acoustic effects, rather than extracting or reasoning over structured financial data.

\paragraph{Medical QA} AI models are increasingly utilized across medical field, tackling a wide array of applications such as diagnostic assistance through analysis of medical images (e.g., X-rays, MRIs) \cite{wei2023deep, wang2024robust, wei2025hierarchical} and dialogues \cite{varshney2023knowledge, li2024distinct}, drug discovery \cite{dara2022machine, blanco2023role}, digital biomarkers \cite{arya2023convergence, zhou2024glumarker}, and personalized patient care \cite{kasula2023harnessing, li2023bilateral}.
Among these, Medical Question Answering (Medical QA) \cite{abacha2015means, goodwin2016medical} is a specialized field focused on developing systems that understand and respond to health-related queries. 
These systems often process information from diverse sources to provide accurate medical information or support clinical decision-making. In contrast, Music AVQA centers on interpreting music-related videos, requiring models to reason over dense, continuous audio and tightly synchronized visual streams. While both involve multimodal and complex reasoning, Music AVQA uniquely demands fine-grained alignment of perceptual cues in expressive performance contexts, such as gesture, rhythm, and phrasing.

\clearpage

\end{document}